\documentclass[twocolumn,american,aps,onecolumn]{revtex4}
\usepackage[T1]{fontenc}
\usepackage[latin9]{inputenc}
\usepackage{amsmath}
\usepackage{amssymb}
\usepackage{graphicx}

\makeatletter

\newcommand{\noun}[1]{\textsc{#1}}

\@ifundefined{textcolor}{}
{%
 \definecolor{BLACK}{gray}{0}
 \definecolor{WHITE}{gray}{1}
 \definecolor{RED}{rgb}{1,0,0}
 \definecolor{GREEN}{rgb}{0,1,0}
 \definecolor{BLUE}{rgb}{0,0,1}
 \definecolor{CYAN}{cmyk}{1,0,0,0}
 \definecolor{MAGENTA}{cmyk}{0,1,0,0}
 \definecolor{YELLOW}{cmyk}{0,0,1,0}
}

\makeatother

\usepackage{babel}
\begin{document}

\title{A topological model for inflation }

\author{Torsten Asselmeyer-Maluga}

\email{torsten.asselmeyer-maluga@dlr.de}

\selectlanguage{american}%

\address{German Aero space Center (DLR), Rosa-Luxemburg-Str 2, 10178 Berlin,
Germany and Copernicus Center for Interdisciplinary Studies, ulica
Szczepa{\'{n}}ska 1/5, 31-011 Krak{ó}w, Poland}

\author{Jerzy Kr\'ol}

\email{iriking@wp.pl}

\selectlanguage{american}%

\address{University of Silesia, Institute of Physics, ul. Uniwesytecka 4,
40-007 Katowice, Poland and Copernicus Center for Interdisciplinary
Studies, ulica Szczepa{\'{n}}ska 1/5, 31-011 Krak{ó}w, Poland}
\begin{abstract}
In this paper we will discuss a new model for inflation based on topological
ideas. For that purpose we will consider the change of the topology
of the spatial component seen as compact 3-manifold. We analyzed the
topology change by using Morse theory and handle body decomposition
of manifolds. For the general case of a topology change of a $n-$manifold,
we are forced to introduce a scalar field with quadratic potential
or double well potential. Unfortunately these cases are ruled out
by the CMB results of the Planck misssion. In case of 3-manifolds
there is another possibility which uses deep results in differential
topology of 4-manifolds. With the help of these results we will show
that in case of a fixed homology of the 3-manifolds one will obtain
a scalar field potential which is conformally equivalent to the Starobinsky
model. The free parameter of the Starobinsky model can be expressed
by the topological invariants of the 3-manifold. Furthermore we are
able to express the number of e-folds as well as the energy and length
scale by the Chern-Simons invariant of the final 3-manifold. We will
apply these result to a specific model which was used by us to discuss
the appearance of the cosmological constant with an experimentally
confirmed value.
\end{abstract}



\maketitle
\tableofcontents{}

\section{Introduction}

Because of the influx of observational data%
\footnote{In particular, see the recent results of the Planck satellite in arXiv
from 1303.5062 to 1303.5090.%
}, recent years have witnessed enormous advances in our understanding
of the early universe. To interpret the present data, it is sufficient
to work in a regime in which spacetime can be taken to be a smooth
continuum as in general relativity, setting aside fundamental questions
involving the deep Planck regime. However, for a complete conceptual
understanding as well as interpretation of the future, more refined
data, these long-standing issues will have to be faced squarely. As
an example one may ask, can one show from first principles that the
smooth spacetime of general relativity is valid at the onset of inflation?
At the same time, this approach has some problems like special initial
conditions and free parameters like the amount of increase (number
of e-folds). Furthermore, there are many possibilities for a specific
model (chaotic or fractal inflation, Starobinsky model etc.). But
the impressive results of the PLANCK mission excludes many models
\cite{PlanckInflation2013}. Nevertheless, the main questions for
inflation remain: what is the scalar field? what is the number of
e-folds? which model is realistic? what is the energy scale? etc.
In this paper we will focus mainly on the question about the origin
of inflation. Today inflation is the main theoretical framework that
describes the early Universe and that can account for the present
observational data \cite{WMAP-7-years,PlanckCosmoParameters2013,PlanckCosmoParameters2015}.
In thirty years of existence \cite{Guth1981,Linde1982}, inflation
has survived, in contrast with earlier competitors, the tremendous
improvement of cosmological data. In particular, the fluctuations
of the Cosmic Microwave Background (CMB) had not yet been measured
when inflation was invented, whereas they give us today a remarkable
picture of the cosmological perturbations in the early Universe. In
nearly all known models, the inflation period is caused by one or
more scalar field(s)\cite{InflationBook}. But the question about
the origin of this scalar field remains among other problems (see
for instance \cite{Penrose1989}).

In this paper we will go a different way to explain inflation. Inspired
by Wheelers idea of topology change at small scales in quantum gravity,
we will consider a spatial topology change. The description of the
change using the concept of a cobordism (representing the spacetime)
will lead automatically to a scalar field. On general grounds, one
can show that there are two kinds of changes: adding a submanifold
or change/deform a submanifold. In the first case, we will get the
quadratic potential of chaotic inflation whereas in the second case
we will obtain the double well potential of topological inflation.
But both models were ruled out by the PLANCK mission. Amazingly, only
in four dimension there is another possibility of a topology change.
This modification is an infinite process of submanifold deformations
(arranged along a tree). At the first view, it seems hopeless to calculate
something. But in contrast to the two cases above, we are able to
determine everything in this model. The potential of the scalar field
$\phi$ is $(1-e^{-\phi})^{2}$ which is conformally equivalent to
the Starobinsky model. The number of e-folds is determined by a topological
invariant of the spatial space. This invariant will be used to get
expressions for $\alpha$ (free parameter in the Starobinsky model)
or the energy scale of inflation. Why is this miracle possible? Mostow-Prasad
rigidity (see \ref{sec:Appendix-Hyperbolic-Mostow}) is the cause
for this behavior. The infinite process of submanifold deformations
implies a hyperbolic geometry for the underlying space. But any deformation
of a hyperbolic space must be an isometry. Therefore geometric expressions
like volume or curvature are topological invariants. It is the point
where geometry and topology meet. In a previous paper \cite{AsselmeyerKrol2018a}
we discussed a concrete model of topology change in the evolution
of the cosmos with two phases. In particular, we obtained a realistic
value of the cosmological constant. Here we will use this model to
calculate the values of the inflation parameters in the Starobinsky
model, i.e. $\alpha$ (coupling of the $R^{2}$ term), energy scale,
number of e-folds $N$, the spectral tilt ${\displaystyle n_{s}}$
and the tensor-scalar ratio $r$.  We will also compare these values
with the current measurements. One point remains, how does this model
couples to matter (reheating)? In a geometric/topological theory of
inflation one also needs a geometric model of matter to explain this
coupling. Fortunately, this theory was partly developed in previous
work \cite{AsselmeyerRose2012,AsselmeyerBrans2015}. By using these
ideas, we will explain the coupling between the scalar field and matter.
The coupling constant is given by a topological invariant again.

\section{The Model}

The main idea of our model can be summarized by a simple assumption:
\emph{during the cosmic evolution (i.e. directly after the Big Bang)
the spatial component (space) undergoes a topology change}. This assumption
is mainly motivated by all approaches to quantum gravity. Notable
are first ideas by Wheeler \cite{Wheeler62}. But topology changes
are able to produce singularities and causal discontinuities as shown
in \cite{deWitteAnderson1986}. In many cases one can circumvent these
problems as discussed in \cite{Dowker1997,DowkerGarcia:1998,DowkerGarciaSurya:2000}.
Here, we will implicitly assume that the topology changes is causal
continuous.

At first we have to discuss the description of a spatial topology
change. Let $\Sigma_{1}$ and $\Sigma_{2}$ be two compact closed
3-manifolds so that $\Sigma_{1}$ is changed to $\Sigma_{2}$. Now
there is a spacetime $M$ with $\partial M=\Sigma_{1}\sqcup\Sigma_{2}$
called a cobordism. For now we have to face the question to characterize
the topology of the cobordism. It is obvious that two diffeomorphic
3-manifolds $\Sigma_{1}=\Sigma_{2}$ will generate a trivial cobordism
$\Sigma_{1}\times[0,1]$. Interestingly, it is not true if there is
a counterexample to the smooth Poincare conjecture in dimension 4.
Then this cobordism between diffeomorphic 3-manifold can be also non-trivial
(i.e. non-product). In general, the difference between two manifolds
is expressed in a complicated topological structure of the interior.
The prominent example is a cobordism between two disjoint circles
and one circle, the so-called trouser. There, the non-triviality of
the cobordism is given by the appearance of a 'singular' point, the
crotch. Fortunately, this behavior can be generalized to all other
cases too. To understand this solution we have to introduce Morse
theory and handlebody decomposition of manifolds. By using these methods
we will show that a topology change requires a scalar field including
an interaction potential (related to the so-called Morse function).

\subsection{Morse theory and handles}

In Morse theory one analyzed the (differential-)topology of a manifold
$M$ by using a (twice-)differentiable function $f:M\to\mathbb{R}$.
The main idea is the usage of this function to generate a diffeomorphism
via the gradient equation 
\begin{equation}
\frac{d}{dt}\vec{x}=-\nabla f(\vec{x})\label{eq:Morse-flow}
\end{equation}
in a coordinate system. Away from the fix point $\nabla f=0$, the
solution of this differential equation is the desired diffeomorphism.
This behavior breaks down at the fix points. The fix points of this
equation are the critical points of $f$. Now one has to assume that
these critical points are isolated and that the matrix of second derivatives
has maximal rank (non-degenerated critical points). This function
$f$ is called a Morse function. Then, the function $f$ in a neighborhood
of an isolated, non-degenerated point $x^{(0)}$ (i.e. $\nabla f|_{x^{(0)}}=0$,
$\det\left(\frac{\partial^{2}f}{\partial x_{i}\partial x_{j}}|_{x^{(0)}}\right)\not=0$)
looks generically like
\begin{equation}
f(x)=f(x^{(0)})-x_{1}^{2}-x_{2}^{2}-\cdots-x_{k}^{2}+x_{k+1}^{2}+\cdots+x_{n}^{2}\label{eq:Morse-function}
\end{equation}
where the number $k$ is called the index of the critical point. In
the physics point of view, the Morse function is a scalar field over
the manifold. The Morse function (\ref{eq:Morse-function}) at the
critical point is a quadratic form w.r.t the coordinates. This form
is invariant by the action of the group $SO(n-k,k)$. If one interpret
this group as isometry group then one can determine the geometry in
the neighborhood of the critical point. But the the Morse function
reflects the topological properties of the underlying space, i.e.
the analytic properties of $f$ are connected with the topology of
$M$. For that purpose we will define the level set of $f$, i.e.
\[
M(a)=f^{-1}((-\infty,a])=\left\{ x\in M|\: f(x)\leq a\right\} 
\]
Now consider two sets $M(a)$ and $M(b)$ for $a<b$. If there is
no critical point in the compact set $f^{-1}([a,b])$ then $M(a)$
and $M(b)$ are homotopy equivalent (and also topologically equivalent,
at least in dimension smaller than 5). If the compact set $f^{-1}([a,b])$
contains a critical point of index $k$ then $M(a)$ and $M(b)$ are
related by the attachment of a $k-$handle $D^{k}\times D^{n-k}$
($D^{k}=\left\{ x\in\mathbb{R}^{k+1}|\:||x||^{2}\leq1\right\} $ is
the $k-$disc), i.e. $M(b)=M(a)\cup D^{k}\times D^{n-k}$. Therefore,
the topology of $M$ is encoded in the critical values of the Morse
function. But let us give two words of warning: firstly this approach
gives only the number handles but not the detailed attachment of the
handles and secondly the number of handles as induced by the Morse
function can be larger as the minimal number of handles used to decompose
the manifold. Both facts are expressed in the Morse relations: let
$n_{k}$ be the number of critical points of index $k$ and $b_{k}$
the $k$th Betti number (= the rank of the homology group with values
in $\mathbb{R}$). Then, one has
\[
b_{k}\leq n_{k}\qquad\sum_{k}(-1)^{k}n_{k}=\sum_{k}(-1)^{k}b_{k}=\chi(M)
\]
where $\chi(M)$ is the Euler characteristics of $M$. So, Morse theory
extracted only the homological properties of the manifold but not
the whole topological information.

\subsection{Cobordism, handles and Cerf theory}

\begin{figure}
\includegraphics[scale=0.25]{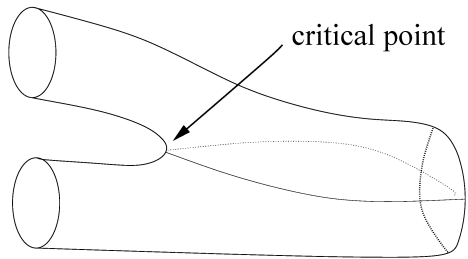}

\caption{a pant as cobordism \label{fig:a-pant-as-cobordism}}

\end{figure}

Now we will discuss the cobordism $W(\Sigma_{1},\Sigma_{2})$ between
two different 3-manifolds $\Sigma_{1}$ and $\Sigma_{2}$. In case
of 3-manifolds, the word 'different' means non-diffeomorphic which
agrees with non-homeomorphic (see \cite{Moi:52}). What is the structure
of the cobordism $W(\Sigma_{1},\Sigma_{2})$ for different 3-manifolds?
The answer can be simply expressed that there are one or more $k-$handles
in the interior of the cobordism. A proof can be found in \cite{Mil:65}
and we will discuss a simple example now. Let us consider a pant or
trouser like above, i.e. a cobordism between two disjoint circles
and one circle (see Fig. \ref{fig:a-pant-as-cobordism}). 
 A circle is the union of one 0-handle ($D^{0}\times D^{1}$) and
one 1-handle ($D^{1}\times D^{0}$) glued together along $\partial D^{1}\times D^{0}=S^{0}\times D^{0}$
and the boundary of the 0-handle $\partial(D^{0}\times D^{1})=S^{0}$,
i.e. along the end-points of the two intervals $D^{1}$. Now lets
go from two disjoint circles on one side of the cobordism to one circle
of the other side of the cobordism. The two disjoint circles are build
from two 0-handles and two 1-handles whereas the one circle is decomposed
by one 0-handle and one 1-handle. Therefore the pair of one 0-handle
and one 1-handle was destroyed. Each process of this kind will produce
a handle in the interior of the cobordism. In this case it is a 1-handle
(the crotch of the pant). The critical point of this handle represents
the topology change. It is the critical point of the corresponding
Morse function for the cobordism. Of course the process can be reversed
(cobordism classes are forming a group). Then a 0-/1-handle pair appears
and one circles splits into two disjoint circles but the main observation
is the same: A topology change produces an additional handle in the
interior of the cobordism. For the discussion later, we have to make
an important remark. The 1-handle in the interior of the cobordism
is a saddle and the critical point is a saddle point. The corresponding
Morse function is given by $f(x_{1},x_{2})=x_{2}^{2}-x_{1}^{2}$.
Now we are interested in the geometry of the cobordism. The group
$SO(1,1)$ fixes the function $f(x_{1},x_{2})$ by the usual action.
The group $SO(1,1)$ is the group of hyperbolic rotations preserving
the area and orientation of a unit hyperbola. It is a subgroup $SO(1,1)\subset SL(2,\mathbb{R})$
of the Möbius group, the isometry group of the 2-dimensional hyperbolic
space. Because of the saddle point, the interior of the cobordism
is a saddle surface having a hyperbolic geometry (with negative curvature).
This observation can be generalized to any saddle point of a cobordism
(or to any $k-$handle for $0<k<n$ with the dimension $n$ of the
cobordism). In dimension four, one obtains hyperbolic geometries for
1- and 3-handles and the geometry $AdS_{3}\times S^{1}$ for 2-handles
\cite{so2-2-symmetry} for the isometry group $SO(2,2)$ inside of
the cobordism. Then two geodesics will be separated exponentially
after passing the critical point of the handle. This behavior explains
also the appearance of an inflationary phase after a topology change
which will be discussed later.

\begin{figure}
\includegraphics{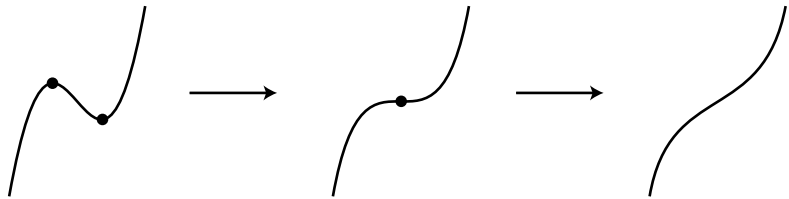}

\caption{Killing a 0- and a 1-handle which can be described by the function
$x^{3}-t\cdot x$ from $t=1$ over $t=0$ to $t=-1$ (from left to
right)\label{fig:killling-handles}}
\end{figure}

The process of 'killing' the 0-/1-handle pair is visualized in Fig.
\ref{fig:killling-handles}. 
 There, one can also find an analytic expression for this process.
For completeness, we remark that the theory behind this description
is called Cerf theory \cite{Cer:70}. In general, the modification
of any handle structure can be simplified to one process of this kind.
The idea of Cerf theory can be simply expressed by considering the
function $W\to\mathbb{R}$, i.e. a one-parameter family of Morse functions
at the boundary of the cobordism. Central point of Cerf theory is
the existence of two generic singularities (with vanishing first derivatives).
The first kind is the Morse singularity: $\pm x^{2}$ and the two
other cases $\pm x^{3}$ and $\pm x^{4}$. For these two cases $x^{3},x^{4}$,
there are resolutions as one-parameter families: $x^{3}-tx$ and $x^{4}-tx^{2}$
(with the parameter $t\in\mathbb{R}$). Interestingly, the $x^{4}-$case
can be reduced to the $x^{3}-$case. The one-parameter family $x^{3}-tx$
is visualized in Fig. \ref{fig:killling-handles} for $t=1,0,-1$
(from left to right). Now the cancellation of a $k-/(k+1)-$handle
pair is described by the function
\begin{equation}
f(x)=f(x^{(0)})-x_{1}^{2}-x_{2}^{2}-\cdots-x_{k}^{2}+(x_{k+1}^{3}-t\cdot x_{k+1})+\cdots+x_{n}^{2}\label{eq:Cerf-theory-function}
\end{equation}
and we will use this function to describe the change in topology as
the effect of handle canceling.

\begin{figure}
\includegraphics[scale=0.25]{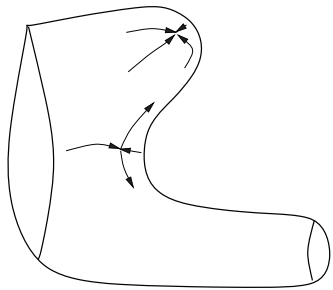}

\caption{cobordism with canceling handle pair \label{fig:cobordism-with-cancelling-pair}}

\end{figure}

One example is the simplification of a handle decomposition. As explained
in the previous subsection, this handle decomposition of the cobordism
can be non-uniquely given. An example is the following picture Fig.
\ref{fig:cobordism-with-cancelling-pair}. 
 A cobordism between two circles (usually the trivial cylinder $S^{1}\times[0,1]$)
is decomposed by an extra pair of one 1-handle and one 2-handle. But
as the figure indicated, there is a flow (determined by the corresponding
Morse function, see (\ref{eq:Morse-flow})) from one critical point
(1-handle) to the other critical point (2-handle). This flow is a
diffeomorphism which can be used to cancel both critical points, see
\cite{Mil:63} and \cite{Mil:65} for the details. For the successful
canceling, one needs an implicit assumption which will be explained
later. 

There is another example where canceling pairs appear, the so-called
homology cobordism. This example will become important later. It is
motivated by the following question: how does a simple-connected,
4-dimensional cobordism with trivial homology look like? In general
one would expect that the corresponding 3-manifolds have to be simply
connected too. But let us consider a disk $D^{2}$ with boundary $S^{1}$.
$D^{2}$ is simply connected in contrast to the boundary. So, every
non-contractable curve in a 3-manifold can be transformed to a contractable
curve inside of the cobordism by attaching a disk or better a 2-handle
$D^{2}\times D^{2}$. The corresponding change of the cobordism can
be changed by the attachment of a 3-handle which cancels the 2-handle.
A simple argument using the Mayer-Vietoris sequence shows that a cobordism
$W(\Sigma_{1},\Sigma_{2})$ which looks like $S^{3}\times[0,1]$ is
a cobordism between two homology 3-spheres $\Sigma_{1},\Sigma_{2}$,
i.e. 3-manifolds with the same homology like the 3-sphere. One call
this cobordism, a homology cobordism. A sequence of these homology
cobordism will look like $S^{3}\times\mathbb{R}$ but inside of this
cobordism you have an ongoing topology change. Later on we will construct
the Starobinsky model from this cobordism.

\subsection{The physics view on cobordism}

In \cite{Wit:82a}, Witten presented a physics view on Morse theory
using supersymmetric quantum mechanics. This work was cited by many
followers and it was the beginning of the field of topological quantum
field theory. Part of this work will be used to describe the cobordism
and its handle decomposition. As explained above, a cobordism $W(\Sigma_{1},\Sigma_{2})$
between different manifolds$\Sigma_{1}$ and $\Sigma_{2}$ contains
at least one handle, say $k-$handle $H_{k}$. This handle $H_{k}$
is embedded in the cobordism by using a map
\[
\Phi_{H}:H_{k}\hookrightarrow W(\Sigma_{1},\Sigma_{2})
\]
to visualizing the attachment of the handle. It can be locally modeled
by a map
\[
\Phi_{H,loc}:H_{k}\hookrightarrow U\subset\mathbb{R}^{4}
\]
where $U\subset W(\Sigma_{1},\Sigma_{2})$ is a chart of the cobordism
representing the attachment of the handle. In principle, this local
description is enough to understand the adding of a handle to the
interior of the cobordism (and representing the non-triviality of
the cobordism). So, if we are choosing the map 
\[
\Phi:W(\Sigma_{1},\Sigma_{2})\to\mathbb{R}^{4},\qquad supp(\Phi)=H_{k}
\]
then we have an equivalent description given by a set of four scalar
fields $(\phi_{0},\ldots,\phi_{3})=\Phi$. Importantly, this description
can be generalized to all dimensions expressing the topology change.
In the special case of 3-manifolds we will later present another description
using a $SU(2)-$valued scalar field. 

But now we will consider the case of a $n-$manifold. It is not an
accident that scalar fields are describing a topology change. Critical
points of a scalar field (the Morse function) express the topology
of the underlying space, at least partly. Here, in the case of a topology
change we have to consider the change of a scalar field. This change
leads to the appearance of one or more handles in the interior of
the cobordism whose location is a tupel of four scalar fields to describe
the embedding. A vector field is not suitable because there is no
'direction' in an embedding. But we are able to simplify this description.
The embedding can be chosen in such a manner that the flow to the
critical point of the handle $H_{k}$ is normal to the boundary. The
Morse function for the handle $H_{k}$ is given by $h(\Phi)=\sum_{i}(\pm\phi_{i}^{2})$
in the coordinate system $(\phi_{1},\ldots,\phi_{n})$. The normal
direction in the cobordism is expressed by a coordinate $t$ seen
as a (locally) non-vanishing vector field (as section of the normal
bundle of the boundary $\partial W(\Sigma_{1},\Sigma_{2})$). The
case of a cobordism with more than one critical point (or handle)
is more interesting. This case includes also the situation to simplify
the cobordism (see Fig. \ref{fig:cobordism-with-cancelling-pair}).
In\cite{Wit:82a}, this situation was considered. The Morse function
$h$ inside of the cobordism generates the 'potential energy' of the
problem to be $V(\Phi)=(dh)^{2}$. Then the flow from one critical
point to another critical point can be described as a tunneling path.
These paths are the paths of steepest descent (leading from one critical
point to another critical point) expressed as solutions of the equation
(\ref{eq:Morse-flow}) now written as
\[
\frac{d\phi_{i}}{dt}=g^{ij}\frac{\partial h}{\partial\phi_{j}}
\]
with respect to a metric $g_{ij}$ of $W(\Sigma_{1},\Sigma_{2})$.
Here, the variable $t$ the path of steepest descent. But as shown
in \cite{Mil:65}, one can order the handles so that the coordinate
$t$ of the cobordism cane be identified with this parameter which
will be done now. As Witten pointed out in the paper \cite{Wit:82a},
the relevant action is given by 
\begin{equation}
S=\intop dt\left[\frac{1}{2}g^{ij}\frac{d\phi_{i}}{dt}\frac{d\phi_{j}}{dt}+\frac{1}{2}g^{ij}\frac{\partial h}{\partial\phi_{i}}\frac{\partial h}{\partial\phi{}_{j}}\right]\label{eq:Witten-Morse-action}
\end{equation}
Now, we identify the coordinates $(x^{0},\ldots,x^{n-1})$ of the
handle $H^{k}$ with one direction in the coordinate system of the
cobordism, say $x^{0}=t$, and using a Lorentz transformation at the
same time then no direction is preferred (a standard argument). Finally
we obtain the action of the nonlinear sigma model
\[
S=\intop\left(\frac{1}{2}g^{kl}\partial_{k}\phi^{i}\partial_{l}\phi_{i}+\frac{1}{2}g^{ij}\frac{\partial h}{\partial\phi_{i}}\frac{\partial h}{\partial\phi{}_{j}}\right)
\]
and the path of steepest descent is given by a choice of a function
$t=t(x^{i})$ (as embedding of the curve). This action can be also
used to describe the canceling process of a $k-/(k+1)-$handle pair.
Both handles agreed in nearly all directions except one direction.
It is the direction $k+1$ with coordinate $\phi_{k+1}$. The $k-$handle
is given by the Morse function $-\phi_{1}^{2}\text{-\ensuremath{\cdots}-}\phi_{k}^{2}+\phi_{k+1}^{2}+\cdots+\phi_{n}^{2}$
whereas the $(k+1)-$handle is determined by the Morse function $-\phi_{1}^{2}\text{-\ensuremath{\cdots}-}\phi_{k+1}^{2}+\phi_{k+2}^{2}+\cdots+\phi_{n}^{2}$.
The difference between both Morse functions is concentrated at the
$(k+1)-$direction: the function $+\phi_{k+1}^{2}$ for the $k-$handle
and $-\phi_{k+1}^{2}$ for the $(k+1)-$handle. Both handles are connected
along this direction and the canceling of both handles has its origin
in this connection. In the above mentioned Cerf theory \cite{Cer:70},
this handle pair is described by one function
\begin{equation}
-\phi_{1}^{2}\text{-\ensuremath{\cdots}-}\phi_{k}^{2}+\left(\phi_{k+1}^{3}-T\cdot\phi_{k+1}\right)+\phi_{k+2}^{2}+\cdots+\phi_{n}^{2}\label{eq:Cerf-function}
\end{equation}
with one parameter $T$. The main result of Cerf theory states that
this expression is unique (or better generic) up to diffeomorphisms.
The canceling is described schematically in Fig. \ref{fig:killling-handles}
for the parameter $T=1$, $T=0$ and $T=-1$. But then we need only
\emph{one scalar field} in the action (\ref{eq:Witten-Morse-action})
and the function $h$ is given by the bracket term in the expression
(\ref{eq:Cerf-function}). Finally we obtain the action
\[
S=\intop dt\left[\frac{1}{2}\frac{d\phi}{dt}\frac{d\phi}{dt}+\frac{1}{2}\frac{\partial h}{\partial\phi}\frac{\partial h}{\partial\phi}\right]=\intop dt\left[\frac{1}{2}\frac{d\phi}{dt}\frac{d\phi}{dt}+\frac{1}{2}\left(\phi^{2}-T\right)^{2}\right]
\]
for $h=\phi^{3}/3-T\cdot\phi$. But as explained above, one has the
freedom to embed the handle pair into the $x-$coordinate system of
the cobordism. For the curve $x(t)$ connecting the two handles, one
can rewrite the total derivative 
\[
\frac{d}{dt}=\dot{x}^{\mu}\partial_{\mu}
\]
and in a small neighborhood we can use the usual relation $\dot{x}^{\mu}\dot{x}_{\mu}=c^{2}$
between the four velocities. Instead to integrate only along the curve,
we will consider a field $\phi$ of handle pairs on the cobordism
$W(\Sigma_{1},\Sigma_{2})$. Then $\phi$ can be interpreted as a
kind of density for handle pairs in $\Sigma_{1,2}$. Then we have
to integrate over the whole cobordism to obtain the action
\[
S=\intop_{W(\Sigma_{1},\Sigma_{2})}d^{n}x\left[\frac{1}{2}\partial_{\mu}\phi\partial^{\mu}\phi+\frac{1}{2}\left(\phi^{2}-T\right)^{2}\right]
\]
where we set $c=1$. The argumentation can be generalized to other
cases like the appearance of a handle as described by the function
$h=\pm\phi^{2}$ leading to the general action
\begin{equation}
S=\intop_{W(\Sigma_{1},\Sigma_{2})}d^{n}x\left[\frac{1}{2}\partial_{\mu}\phi\partial^{\mu}\phi+\frac{1}{2}\left(\frac{dh}{d\phi}\right)^{2}\right]\label{eq:general-cobordism-action}
\end{equation}
As remarked above, we looked for the generic cases like $h=\pm\phi^{2}$
and $h=\phi^{3}-T\phi$ but higher powers are also possible. In combination
with the Einstein-Hilbert action we obtain the two generic models
\begin{eqnarray*}
S_{chaotic} & = & \intop_{W(\Sigma_{1},\Sigma_{2})}d^{n}x\sqrt{g}\left[R+\frac{1}{2}\partial_{\mu}\phi\partial^{\mu}\phi+\frac{1}{2}\phi^{2}\right]\\
S_{topological} & = & \intop_{W(\Sigma_{1},\Sigma_{2})}d^{n}x\sqrt{g}\left[R+\frac{1}{2}\partial_{\mu}\phi\partial^{\mu}\phi+\frac{1}{2}\left(\phi^{2}-T\right)^{2}\right]
\end{eqnarray*}
of chaotic inflation and topological inflation. 

By using the model of a topology change, we are able to reproduce
two known inflationary models. The advantage of this model is the
natural appearance of the scalar field $\phi$ which is associated
to the topology change. Unfortunately, both models were ruled out
by recent results of the Planck mission \cite{PlanckInflation2013}.
Only potentials like $\phi^{m}$ with $1<m<2$ or long-tailed expressions
like $(1-e^{-\phi})^{2}$ are possible. But functions like $\phi^{m}$
cannot be generated by Morse functions and one cannot reproduce this
model by topological methods. Interestingly, only in dimension four
there is the possibility to obtain long-tailed expressions. There
is a simple reason why this is possible: in dimension four a pair
of handles like $1-/2-$or $2-/3-$handle pairs cannot be canceled
smoothly but topologically using an infinite process. In the next
section we will describe this process which will lead to the Starobinsky
model which is one of the favored models of the Planck mission.

\section{Inflation in four dimensions }

In this section we will specialize to 4D spacetime. Here, the cancellation
of a handle pair is described by an infinite process i.e. one needs
a special handle known as Casson handle. Casson handles are parametrized
by all trees. In principle, this fact is the reason for the exponential
potential leading to Starobinsky inflation.

\subsection{Cobordism between 3-manifolds and 4-manifold topology}

\begin{figure}
\includegraphics[scale=4]{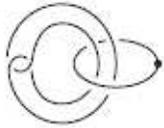}

\caption{Example of non-canceling 1-/2-handle pair, the 1-handle is visualized
by the circle with the dot \label{fig:Example-of-non-cancelling-handle-pair}}

\end{figure}

In the previous section we described the general case of a topology
change. Implicitly we assumed that the canceling of $k-/(k+1)-$handles
is always possible. But in dimension 4, there is a problem which is
at the heart of all problems in 4-dimensional topology. As an example
let us consider a pair of one 2-handle $D^{2}\times D^{2}$ and one
1-handle $D^{1}\times D^{3}$. The 1-handle and 2-handle cancel each
other if the attaching sphere $\partial D^{2}\times0=S^{1}$ of the
2-handle meets the belt sphere $0\times\partial D^{3}=S^{2}$ of the
1-handle tranversally in one point. To understand the problem, we
have to consider the attachment of handles. A $k-$handle $D^{k}\times D^{n-k}$
is attached to $D^{n}$ via the boundary by the map $\partial D^{k}\times D^{n-k}\to\partial D^{n}$.
Then a 2-handle is attached to $\partial D^{4}=S^{3}$ by an embedding
of $\partial D^{2}\times D^{2}=S^{1}\times D^{2}$, the solid torus.
But this map is equivalent to $S^{1}\to S^{3}$, i.e. the attaching
of a 2-handle is determined by a knot. There is also an additional
number, the framing, which describes how a parallel copy of the knots
wind around the knot. The attachment of 1- and 3-handles are easier
to describe. In case of a 1-handle, Akbulut \cite{AkbKir:79} found
another amazing description: a 1-handle is a removed 2-handle with
fixed framing (see \cite{GomSti:1999} section 5.4 for the details).
An example of a non-canceling pair of one 1-handle and one 2-handle
is visualized in Fig. \ref{fig:Example-of-non-cancelling-handle-pair}.
 In this example, the attaching sphere of the 2-handle (the knot without
the dot) meets the belt sphere of the 1-handle (the knot with the
dot) twice. Usually, a curve meeting a second curve twice can be separated.
This process is called the Whitney trick. For the realization of this
process in a controlled manner, one needs an embedded disk. But it
is known that the Whitney trick fails because the disk contains self-intersections
(it is immersed in contrast to embedded), see \cite{Asselmeyer2007}.
Interestingly, it is possible to realize the Whitney trick topologically.
But then one needs an infinite process as shown by Freedman \cite{Fre:82}.
Now we will describe this process to use it for the Starobinsky model.

\subsection{Casson handles or the infinite process of handle-cancellations}

In dimension 4, the process of handle canceling can be an infinite
process. One can understand the reason simply. Two disks in a 4-manifold
intersect in a point. At the same time, disks in a 4-manifold can
admit self-intersections, i.e. we obtain an immersed disk (in contrast
to an embedded disk with no self-intersections). As explained above,
one can cancel the self-intersections but one needs another disk admitting
self-intersections again. As Freedman \cite{Fre:82} showed, one needs
infinitely many disks (or stages) to cancel the self-intersection
topologically. In this process, an immersed disk can admit more than
one self-intersection and therefore needs more than one disk for its
canceling. Thus, one obtains a tree $T_{CH}$ of disks, called a Casson
handle $CH$ (see the appendix and \cite{Cas:73,Fre:79,Fre:82,GomSti:1999}).
Here we have also a special situation: an infinite object, the Casson
handle $CH$, has to be embedded into a compact 4-manifold, the cobordism
$W(\Sigma_{1},\Sigma_{2})$. Furthermore, the tree can be exponentially
large (see \cite{Biz:94}). But then we have to find an embedding
of this tree into a compact submanifold. For simplicity we can assume
a disk $D^{2}$ of radius $1$. The infinite tree has a root and a
continuum of leaves at infinity. If we put the root of the tree in
the middle of the disk then the leaves have to be at the boundary
of the disk. The corresponding disk has to admit a special geometry
to reflect this properties, it is the Poincare hyperbolic disk with
metric
\[
ds_{2D}^{2}=\frac{dx^{2}+dy^{2}}{\left(1-x^{2}-y^{2}\right)^{2}}
\]
The boundary of the disk (i.e. $x^{2}+y^{2}=1$) represents the point
at infinity. The (scalar) curvature is negative, i.e. the disk carries
a hyperbolic metric. This metric can be simply transformed into 
\begin{equation}
ds^{2}=g_{rr}dr^{2}+g_{\xi\xi}d\xi^{2}=\frac{dr^{2}+r^{2}d\xi^{2}}{\left(1-r^{2}\right)^{2}}\label{eq:radial-metric-hyp-disk}
\end{equation}
by using $x=r\cdot\cos\xi$, $y=r\cdot\sin\xi$. The tree of the Casson
handle is embedded along a fixed angle $\xi$, i.e. $d\xi=0$. As
mentioned above, the root of the tree is located at the center of
the disk. Then the whole tree is located between $0<r<1$ where $r=1$
is containing the leaves of the tree. By using $d\xi=0$, we obtained
for the metric
\begin{equation}
ds^{2}|_{tree}=\frac{dr^{2}}{\left(1-r^{2}\right)^{2}}=\left(d\left(\ln\left(\frac{1+r}{1-r}\right)\right)\right)^{2}=d\phi^{2}\label{eq:metric-hyp-disk}
\end{equation}
by choosing
\begin{equation}
\phi=\ln\left(\frac{1+r}{1-r}\right)\,,\qquad r=\frac{e^{\phi}-1}{e^{\phi}+1}\:.\label{eq:defining-scalar-field}
\end{equation}
A Morse function on the disk is given by 
\[
h(r)=\pm r^{2}
\]
and a cancelling pair (by using Cerf theory) can be expressed by
\[
h(r)=r^{3}+T\cdot r
\]
with the deformation parameter $T$. For $T<0$ one has the canceling
pair of two handles and for $T>0$ both handles are disappeared. The
cancellation point is given by $T=0$ or by $h(r)=r^{3}$. For the
following argumentation, we will start with a Morse function $h(r)=r^{2}/2$
and deform it to a pair of two handles by choosing
\begin{equation}
h(r)=\frac{r^{2}}{2}-\frac{r^{3}}{3}\,.\label{eq:Casson-handle-Morse-function}
\end{equation}
One handle is at the middle of the disk ($r=0$) and the canceling
handle is located at infinity (after adding the whole Casson handle),
i.e. at the boundary of the disk $r=1$. One can also construct the
previous Morse function directly from this data. For that purpose
, we have to choose the first derivative to be
\[
dh=r\left(1-r\right)dr
\]
to get critical points at $r=0$ (minimum) and $r=1$ (maximum). By
a simple integration, one will get the Morse function (\ref{eq:Casson-handle-Morse-function})
above. Then in the action (\ref{eq:general-cobordism-action}), one
has the potential 
\[
V(r)=g_{rr}\left(\frac{\partial h}{\partial r}\right)^{2}=\frac{(1-r)^{2}r^{2}}{\left(1-r^{2}\right)^{2}}
\]
with respect to the metric (\ref{eq:radial-metric-hyp-disk}) 
\[
g_{rr}=\frac{1}{(1-r^{2})^{2}}\,.
\]
In our philosophy, we have to use the coordinate $\phi$ instead of
$r$ which is equivalent to transform the problem back into the Euclidean
space. Then we will obtain the scalar field $\phi$ and the new potential
$V(\phi)$ in these coordinates
\begin{equation}
V(\phi)=e^{-2\phi}\left(e^{\phi}-1\right)^{2}=\left(1-e^{-\phi}\right)^{2}\label{eq:Morse-function-CH}
\end{equation}
leading to the action
\begin{equation}
S_{Starobinsky}=\intop_{W(\Sigma_{1},\Sigma_{2})}d^{4}x\sqrt{g}\left(R+\partial^{\mu}\phi\partial_{\mu}\phi-A\cdot\left(1-e^{-\phi}\right)^{2}\right)\label{eq:Starobinsky-scalar-field-model}
\end{equation}
of the Starobinsky model written as scalar field action. Here we take
the opportunity to scale the potential by the free parameter $A$,
i.e. by scaling the function $V(\phi)\to A\cdot V(\phi)$. The classical
Starobinsky model can be constructed after performing a conformal
transformation
\begin{equation}
g\to g'=e^{\phi}g\label{eq:conformal-trafo-metric}
\end{equation}
with 
\begin{equation}
e^{\phi}=1+2\alpha\cdot R\label{eq:conformal-trafo-to-Starobinsky}
\end{equation}
with the scalar curvature $R$ and $\alpha=\frac{1}{8A}$. Then one
obtains
\begin{equation}
S_{Starobinsky}=\intop_{W(\Sigma_{1},\Sigma_{2})}d^{4}x\sqrt{g}\left(R+\alpha\cdot R^{2}\right)\label{eq:Starobinsky-original-model}
\end{equation}
the usual Starobinsky model. It is one of the few models which agrees
with the results of the Planck mission. But what is the meaning of
this conformal transformation? Is it possible to determine the free
parameter $\alpha$ and what is its meaning? What is the real geometric
background of this model? All these question have to be addressed
to get a full derivation of the model. Therefore we will start with
the model (\ref{eq:Starobinsky-scalar-field-model}) to obtain (\ref{eq:Starobinsky-original-model}).

\subsection{A geometric interpretation of the Starobinsky model\label{sub:A-geometric-interpretation}}

The infinite process of handle-cancellation in dimension four was
used to construct the scalar field model (\ref{eq:Starobinsky-scalar-field-model})
which is conformally equivalent to the Starobinsky model. Before we
start we have to give an overview about the model leading to this
action. We considered a topology change of a 3-manifold $\Sigma_{1}$
into another 3-manifold $\Sigma_{2}$ represented by a cobordism,
i.e. by a 4-manifold with boundary $\Sigma_{1}\sqcup\Sigma_{2}$.
In this process, one changed the handle structure of $\Sigma_{1}$
into the handle structure of $\Sigma_{2}$. Some handles will be canceled
and some other handles are created. For the special case of 3-manifolds,
one can consider a scalar field $\phi$ where the variation of this
field gives the topology change. Analytically one has to consider
a one parameter family of functions $\phi^{3}-T\cdot\phi$ for this
creation/annihilation process. But in dimension four, there are problems
to realize this process. One needs a complicated infinite tree-like
structure (Casson handle) to manage this process which has to be embedded
into the compact cobordism. The embedding can be realized by using
the hyperbolic metric of the Poincare hyperbolic disk. At the end
we obtained an analytic expression (see the potential (\ref{eq:Morse-function-CH}))
for the corresponding handle structure of this Casson handle. Now
we will understand the geometric origin of this potential. 

For that purpose, we have to consider the potential (\ref{eq:Morse-function-CH}).
We derived it for the Casson handle embedded in the Poincare hyperbolic
disk. A quick look at the defining formula (\ref{eq:defining-scalar-field})
for the scalar field $\phi$ will give us the defining area: $r$
is between $0\leq r<1$ leading to $0\leq\phi<\infty$. But the final
potential told us more: outside of the Poincare hyperbolic disk, the
scalar field can admit negative values and the potential increased
exponentially. At this point we have to remember on the interpretation
of $\phi$: it is directly the deformation of the 3-manifold $\Sigma_{1}$
into $\Sigma_{2}$. Obviously, this deformation will lead to a deformation
of the metric at the 3-manifold as well. Now we will consider the
metric (\ref{eq:radial-metric-hyp-disk}) 
\[
ds^{2}=d\phi^{2}+sinh(\phi)^{2}d\xi^{2}
\]
by using the coordinates $(\phi,\xi)$. Along the $\phi-$coordinate
we have the exponentially crowing tree and along $\xi$ we have also
an exponential increase given by $\exp\phi$ or large positive $\phi$.
For $\phi>0$, one has an exponential increase $e^{\phi}$ of the
metric (see (\ref{eq:metric-hyp-disk})) induced by the hyperbolic
metric used to embed the tree of the Casson handle. 

The embedding of the tree will mimic also the embedding of the whole
Casson handle. The Casson handle (see Appendix) is homeomorphic to
$D^{2}\times\mathbb{R}^{2}$ (see \cite{Fre:82}) and therefore we
will need a four-dimensional version of the Poincare hyperbolic disk,
the Poincare hyperbolic 4-ball with metric
\begin{equation}
ds_{4D}^{2}=\frac{dr^{2}+r^{2}d\Omega^{2}}{(1-r^{2})^{2}}\label{eq:metric-hyp-4-ball}
\end{equation}
with the angle coordinates $\Omega$ (a tupel of 3 angles) and the
radius $r$. Interestingly the calculation remained the same because
the expression (\ref{eq:metric-hyp-4-ball}) qualitatively agreed
with (\ref{eq:metric-hyp-disk}). The interesting part is independent
of the dimension (see \cite{hyperbolic-geometry}). But there is an
important difference: now we have a four-dimensional hyperbolic submanifold
admitting Mostow rigidity or Mostow-Prasad rigidity. Mostow rigidity
is a powerful property. As shown by Mostow \cite{Mos:68}, every hyperbolic
$n-$manifold $n>2$ with finite volume has this property: \emph{Every
diffeomorphism (especially every conformal transformation) of a hyperbolic
$n-$manifold with finite volume is induced by an isometry.} See \ref{sec:Appendix-Hyperbolic-Mostow}
for more information. Therefore one cannot scale a hyperbolic 3- and
4-manifold with finite volume. The volume $vol(\:)$ and the curvature
(or the Chern-Simons invariant) are topological invariants. Now one
may ask that the embedding of the Casson handle is rather artificial
then generic. But there is a second argument. In the appendix we worked
out how the canceling 1-/2-handle pair with a Casson handle attached
looks like. Especially we will show that the corresponding sequence
of 3-manifolds is a sequence of hyperbolic 3-manifolds of finite volume.
Now we are able to argue similarly: one has
\[
ds_{4D}^{2}=d\phi^{2}+sinh(\phi)^{2}d\Omega^{2}
\]
and one will get an exponential increase by $\exp\phi$ along all
directions. This discussion showed that the scalar field $\phi$ can
be interpreted as the deformation of the cobordism metric via a conformal
transformation
\[
g'=e^{\phi}g
\]
for all positive values $\phi>0$. Which geometrical expression determines
this conformal transformation? The field $\phi$ is directly related
to the radius of the hyperbolic disk via (\ref{eq:defining-scalar-field}).
In a very small neighborhood of $\phi=0$, one has approximately an
Euclidean metric (vanishing curvature $R=0$). For large values $\phi>0$,
the curvature of the curves inside the disk increases (relative to
the background metric). The negative values $\phi<0$ correspond to
the area outside of the hyperbolic disk. Here a curve passing the
hyperbolic disk will be changed according to the negative curvature
of the disk. Then we obtain the simple relation
\[
e^{\phi}=1+f(R)
\]
with the strictly increasing function $f$ (i.e. $f(R)<0$ for $R<0$
and $f(R)>0$ for $R>0$). The simplest function is the linear function,
i.e.
\[
e^{\phi}=1+\epsilon\cdot R
\]
and the positivity of the exponential function implied
\[
R\geq-\frac{1}{\epsilon}\:.
\]
This special conformal transformation will transfer the action (\ref{eq:Starobinsky-scalar-field-model})
to the action (\ref{eq:Starobinsky-original-model})
\[
S_{Starobinsky}=\intop_{W(\Sigma_{1},\Sigma_{2})}d^{4}x\sqrt{g}\left(R+\alpha\cdot R^{2}\right)
\]
with $\epsilon=2\cdot\alpha$. But with the discussion above, we can
interpret the Starobinsky model geometrically. The potential $V(\phi)$
of the scalar action is given by
\[
V(\phi)=\frac{1}{8\alpha}\left(1-e^{-\phi}\right)^{2}
\]
or in terms of the curvature
\[
\frac{1}{8\alpha}\left(1-\frac{1}{1+2\alpha\cdot R}\right)^{2}
\]
This expression is flat for positive curvatures reaching slowly the
value $1/8\alpha$. But for negative values ($-1/2\alpha<R<0$), it
grows rapidly. As discussed above, the positive value is related to
the curvature of a curve in the interior of the hyperbolic disk. The
limit values corresponds to the curvature of the whole disk which
is needed to embeds the whole tree of the Casson handle (see above).
But negative values (or the contraction of the metric) are leading
to a strongly increasing potential or the contraction of the disk
(containing the embedded tree) is impossible. This behavior goes over
to the 4-dimensional case (the embedding of the whole Casson handle
into the Poincare hyperbolic 4-ball). Then we can state:\\
\emph{The Starobinsky model is the simplest realization of Mostow
rigidity, i.e. there is a 4-dimensional hyperbolic submanifold of
curvature $-\frac{1}{2\alpha}$ which cannot be contracted.}\\
This submanifold is also the cause for inflation: it is the reaction
of the incompressibility of the submanifold. But Mostow rigidity has
a great advantage: geometric expression are topological invariants.
For this reason we should be able to determine the free parameter
by the topological invariants of the cobordism.

\subsection{Determine the number of e-folds}

In the usual models of inflation, the number of e-folds $N$ is a
free parameter which will be choose to be $50<N<60$. But topology
in combination with Mostow rigidity should determine this value by
purely topological methods. In \cite{AsselmeyerKrol2014} we described
the way to get this value in principle. Here we will adapt the derivation
of the formula to the case in this paper. For that purpose we we will
state two deep mathematical results, the details can be found in \cite{AsselmeyerKrol2018a}.
Above we introduced a hyperbolic disk (Poincare disk) to embed the
infinite tree. This deep result can be expressed a different manner:
there is no freedom or we have to choose the hyperbolic metric (which
is up to isometries given by (\ref{eq:metric-hyp-disk})). The second
deep result is the representation of the topology change as an infinite
chain of 3-manifolds $Y_{1}\to\cdots\to Y_{\infty}$ where each spatial
3-manifold $Y_{n}$ admits a a homogenous metric of constant negative
curvature. Equivalently this change is given by an infinite chain
of cobordisms 
\[
W_{\infty}=W(Y_{1},Y_{2})\cup_{Y_{2}}W(Y_{2},Y_{3})\cup\cdots
\]
representing the chain of changes. This chain of cobordisms $W_{\infty}$
is also embedded in the spacetime. As shown in \cite{AsselmeyerKrol2018a},
$W_{\infty}$ is a model for an end of an exotic $\mathbb{R}^{4}$or
a model of an exotic $S^{3}\times\mathbb{R}$. As shown in this paper,
the embedded $W_{\infty}$ admits a hyperbolic geometry. This hyperbolic
geometry of the cobordism is best expressed by the metric

\begin{equation}
ds^{2}=dt^{2}-a(t)^{2}h_{ik}dx^{i}dx^{k}\label{eq:FRW-metric-1}
\end{equation}
also called the Friedmann-Robertson-Walker metric (FRW metric) with
the scaling function $a(t)$ for the (spatial) 3-manifold (denoted
as $\Sigma$ in the following). As explained above, the spatial 3-manifold
admits (at least for the pieces) a homogenous metric of constant curvature.
Now we have the following situation: each spatial 3-manifold admits
a hyperbolic metric and the whole process (given as infinite chain
$W_{\infty}$ of cobordisms) admits also a hyperbolic metric of constant
negative scalar curvature $R=-4\Lambda=const.<0$ which is realized
by the equation $Ric=-\Lambda g$, i.e. $\Lambda$ is the so-called
cosmological constant. Then one obtains the equation 
\begin{equation}
\left(\frac{\dot{a}}{a}\right)^{2}=\frac{\Lambda}{3}-\frac{k}{a^{2}}\label{eq:FRW-equation}
\end{equation}
having the solutions $a(t)=a_{0}\sqrt{3|k|/\Lambda}\,\sinh(t\sqrt{\Lambda/3})$
for $k<0$, $a(t)=a_{0}\exp(t\sqrt{\Lambda/3})$ for $k=0$ and $a(t)=a_{0}\sqrt{3|k|/\Lambda}\,\cosh(t\sqrt{\Lambda/3})$
for $k>0$ all with exponential behavior. At first we will consider
this equation for constant topology, i.e. for the spacetime $Y_{n}\times[0,1]$.
But as explained above and see \cite{AsselmeyerKrol2018a}, the embedding
and every $Y_{n}$ admits a hyperbolic structure. Now taking Mostow-Prasad
rigidity seriously, the scaling function $a(t)$ must be constant,
or $\dot{a}=0$. Therefore we will get 
\begin{equation}
\Lambda=\frac{3k}{a^{2}}\label{eq:CC-to-3D-curvature}
\end{equation}
by using (\ref{eq:FRW-equation}) for the parts of constant topology.
Formula (\ref{eq:CC-to-3D-curvature}) can be now written in the form
\begin{equation}
\Lambda=\frac{1}{a^{2}}=^{3}R\label{eq:relation-CC-to-curv}
\end{equation}
so that CC is related to the curvature of the 3D space. By using $a_{n}=\sqrt[3]{vol(Y_{n})}$,
we are able to define a scaling parameter for every $Y_{n}$. By Mostow-Prasad
rigidity, $a_{n}$ is also constant, $a_{n}=const.$. But the change
$Y_{n}\to Y_{n+1}$ increases the volumes of $Y_{n}$, $vol(Y_{n+1})>vol(Y_{n})$,
by adding specific 3-manifolds (i.e. complements of the Whitehead
links). Therefore we have the strange situation that the spatial space
changes by the addition of new (topologically non-trivial) spaces.
To illustrate the amount of the change, we have to consider the embedding
directly. It is given by the embedding of the Casson handle $CH$
as represented by the corresponding infinite tree $T_{CH}$. As explained
above, this tree must be embedded into the hyperbolic space. For the
tree, it is enough to use a 2D model, i.e. the hyperbolic space $\mathbb{H}^{2}$.
There are many isometric models of $\mathbb{H}^{2}$(see the appendix
\ref{sec:Appendix-models-hyp-geom} for two models). Above we used
the Poincare disk model but now we will use the half-plane model with
the hyperbolic metric 
\begin{equation}
ds^{2}=\frac{dx^{2}+dy^{2}}{y^{2}}\label{eq:hyp-half-plane}
\end{equation}
to simplify the calculations. The infinite tree must be embedded along
the $y-$axis and we set $dx=0$. The tree $T_{CH}$, as the representative
for the Casson handle, can be seen as metric space instead of a simplicial
tree. In case of a simplicial tree, one is only interested in the
structure given by the number of levels and branches. The tree $T_{CH}$
as a metric space (so-called $\mathbb{R}-$tree) has the property
that any two points are joined by a unique arc isometric to an interval
in $\mathbb{R}$. Then the embedding of $T_{CH}$ is given by the
identification of the coordinate $y$ with the coordinate of the tree
$a_{T}$ representing the distance from the root. This coordinate
is a real number and we can build the new distance function after
the embedding as 
\[
ds_{T}^{2}=\frac{da_{T}^{2}}{a_{T}^{2}}\quad.
\]
But as discussed above, the tree $T_{CH}$ grows with respect to a
time parameter so that we need to introduce an independent time scale
$t$. From the physics point of view, the time scale describes the
partition of the tree into slices. This main ideas was used in \cite{AsselmeyerKrol2018a}
to get the relation between the number of e-folds $N$ and a topological
invariant (Chern-Simons invariant) of the 3-manifold (as result of
the change). In the following we will describe only the main points
in the derivation of the formula (see \cite{AsselmeyerKrol2018a}
for the details):
\begin{itemize}
\item The growing $ds_{T}^{2}$ of the tree with respect to the hyperbolic
structure is given by 
\[
ds_{T}^{2}=\frac{da_{T}^{2}}{a_{T^{2}}}=d\left(\frac{t}{L}\right)^{2}
\]
This equation agrees with the Friedman equation for a (flat) deSitter
space, i.e. the current model of our universe with a CC. This equation
can be formally integrated yielding the expression 
\begin{equation}
a_{T}(t,L)=a_{0}\cdot\exp\left(\frac{t}{L}\right)\label{eq:integration-general-Friedmann}
\end{equation}

\item One important invariant of a cobordism is the signature $\sigma(W)$,
i.e. the number of positive minus the number of negative eigenvalues
of the intersection form. Using the Hirzebruch signature theorem,
it is given by the first Pontryagin class 
\[
\sigma(W(\Sigma_{1},\Sigma_{2}))=\frac{1}{3}\intop_{W(\Sigma_{1},\Sigma_{2})}tr(R\wedge R)
\]
with the curvature 2-form $R$ of the tangent bundle $TW$. By Stokes
theorem, this expression is given by the difference 
\begin{equation}
\sigma\left(W(\Sigma_{1},\Sigma_{2})\right)=\frac{1}{3}CS(\Sigma_{2})-\frac{1}{3}CS(\Sigma_{1})\label{eq:signature-CS-inv-2}
\end{equation}
of two boundary integrals where
\[
\intop_{\Sigma}tr\left(A\wedge dA+\frac{2}{3}A\wedge A\wedge A\right)=8\pi^{2}CS(\Sigma)
\]
is known as Chern-Simons invariant of a 3-manifold $\Sigma$.
\item Using ideas of Witten \cite{Wit:89.2,Wit:89.3,Wit:91.2} we will interpret
the connection $A$ as $ISO(2,1)$ connection. Note that $ISO(2,1)$
is the Lorentz group $SO(3,1)$ by Wigner-In{ö}n{ü} contraction
or the isometry group of the hyperbolic geometry. For that purpose
we choose 
\begin{equation}
A_{i}=\frac{1}{\ell}e_{i}^{a}P_{a}+\omega_{i}^{a}J_{a}\label{eq:Cartan-connection-1}
\end{equation}
with the length $\ell$ and 1-form $A=A_{i}dx^{i}$ with values in
the Lie algebra $ISO(2,1)$ so that the generators $P_{a},J_{a}$
fulfill the commutation relations 
\[
[J_{a},J_{b}]=\epsilon_{abc}J^{c}\qquad[P_{a},P_{b}]=0\qquad[J_{a},P_{b}]=\epsilon_{abc}P^{c}
\]
with pairings $\langle J_{a},P_{b}\rangle=Tr(J_{a}P_{a})=\delta_{ab}$,
$\langle J_{a},J_{b}\rangle=0=\langle P_{a},P_{b}\rangle$. This choice
was discussed in \cite{Wise2010} in the context of Cartan geometry.
\item For vanishing torsion $T=0$, we obtain 
\[
\intop_{\Sigma}tr(A\wedge F)=\frac{1}{\ell}\intop_{\Sigma}\,^{3}R\sqrt{h}d^{3}x
\]
and finally the relation 
\begin{equation}
8\pi^{2}\cdot\ell\cdot CS(\Sigma)=\frac{3}{2}\intop_{\Sigma}\,^{3}R\sqrt{h}d^{3}x\,.\label{eq:CS-to-EH-2}
\end{equation}

\item From (\ref{eq:signature-CS-inv-2}) it follows that 
\begin{eqnarray*}
\sigma\left(W(\Sigma_{1},\Sigma_{k})\right)=\frac{1}{3}CS(\Sigma_{k})-\frac{1}{3}CS(\Sigma_{k-1})+\frac{1}{3}CS(\Sigma_{k-1})-...-\frac{1}{3}CS(\Sigma_{1})=\\
\frac{1}{3}CS(\Sigma_{k})-\frac{1}{3}CS(\Sigma_{1}).
\end{eqnarray*}

\item Then we identify $t=\ell$ with the time and using (\ref{eq:CS-to-EH-2})
we will obtain the expression 
\begin{equation}
t\cdot CS(\Sigma_{2})=\frac{3}{2}\intop_{\Sigma_{2}}\,^{3}R_{ren}\sqrt{h}d^{3}x\label{eq:scalar-curvature-CS-inv-2}
\end{equation}
where the extra factor $8\pi^{2}$ (equals $4\cdot vol(S^{3})$) is
the normalization of the curvature integral. This normalization of
the curvature changes the absolute value of the curvature into
\begin{equation}
|^{3}R_{ren}|=\frac{1}{8\pi^{2}L^{2}}\label{eq:renormalized-curvature}
\end{equation}
and we choose the scaling factor by the relation to the volume $L=\sqrt[3]{vol(\Sigma_{2})/(8\pi^{2})}$. 
\item Then we will obtain formally 
\begin{equation}
\intop_{\Sigma_{2}}\,|^{3}R_{ren}|\sqrt{h}\, d^{3}x=\intop_{\Sigma_{2}}\frac{1}{8\pi^{2}L^{2}}\sqrt{h}\, d^{3}x=L^{3}\cdot\frac{1}{L^{2}}=L\label{eq:CS-integral-relation-2}
\end{equation}
by using 
\[
L^{3}=\frac{vol(\Sigma_{2})}{8\pi^{2}}=\frac{1}{8\pi^{2}}\intop_{\Sigma_{2}}\sqrt{h}\, d^{3}x
\]
in agreement with the normalization above. Let us note that Mostow-Prasad
rigidity enforces us to choose a rescaled formula 
\[
vol_{hyp}(\Sigma_{2})\cdot L^{3}=\frac{1}{8\pi^{2}}\intop_{\Sigma_{2}}\sqrt{h}d^{3}x\,,
\]
with the hyperbolic volume (as a topological invariant). The volume
of all other 3-manifolds can be arbitrarily scaled. In case of hyperbolic
3-manifolds, the scalar curvature $^{3}R<0$ is negative but above
we used the absolute value $|^{3}R|$ in the calculation. Therefore
we have to modify (\ref{eq:scalar-curvature-CS-inv-2}), i.e. we have
to use the absolute value of the curvature $|^{3}R|$ and of the Chern-Simons
invariant $|CS(\Sigma_{2})|$. By (\ref{eq:scalar-curvature-CS-inv-2})
and (\ref{eq:CS-integral-relation-2}) using 
\[
\frac{t}{L}=\begin{cases}
\frac{3}{2\cdot CS(\Sigma_{2})} & \Sigma_{2}\mbox{ non-hyperbolic 3-manifold}\\
\frac{3\cdot vol_{hyp}(\Sigma_{2})}{2\cdot|CS(\Sigma_{2})|} & \Sigma_{2}\mbox{ hyperbolic 3-manifold}
\end{cases}
\]
a simple integration (\ref{eq:integration-general-Friedmann}) gives
the following exponential behavior 
\[
a(t)=a_{0}\cdot e^{t/L}=\begin{cases}
a_{0}\cdot\exp\left(\frac{3}{2\cdot CS(\Sigma_{2})}\right) & \Sigma_{2}\mbox{ non-hyperbolic 3-manifold}\\
a_{0}\cdot\exp\left(\frac{3\cdot vol_{hyp}(\Sigma_{2})}{2\cdot|CS(\Sigma_{2})|}\right) & \Sigma_{2}\mbox{ hyperbolic 3-manifold}\,.
\end{cases}
\]

\end{itemize}
For the following, we will introduce the shortening 
\begin{equation}
\vartheta=\begin{cases}
\frac{3}{2\cdot CS(\Sigma_{2})} & \Sigma_{2}\mbox{ non-hyperbolic 3-manifold}\\
\frac{3\cdot vol_{hyp}(\Sigma_{2})}{2\cdot|CS(\Sigma_{2})|} & \Sigma_{2}\mbox{ hyperbolic 3-manifold}
\end{cases}\label{eq:shortening-vartheta}
\end{equation}
In principle, the value $\vartheta$ is the number of e-folds but
above (\ref{eq:renormalized-curvature}) we used another normalization
of the curvature. But the curvature is related to $a(t)$ by $1/a^{2}$.
Therefore we have to correct the number of e-folds by the logarithm
$ln(8\pi^{2})$ of the normalization and we will obtain
\begin{equation}
N=\vartheta+ln(8\pi^{2})\label{eq:number-of-e-folds}
\end{equation}
or 
\[
N=\begin{cases}
\frac{3}{2\cdot CS(\Sigma_{2})}+ln(8\pi^{2}) & \Sigma_{2}\mbox{ non-hyperbolic 3-manifold}\\
\frac{3\cdot vol_{hyp}(\Sigma_{2})}{2\cdot|CS(\Sigma_{2})|}+ln(8\pi^{2}) & \Sigma_{2}\mbox{ hyperbolic 3-manifold}
\end{cases}
\]
for the number e-folds. This result determines the number of e-folds
and connect it with topological information of the final 3-manifold.
Interestingly, this result is independent of the embedding and of
the particular Casson handle. As shown in the following, using this
result we are able to determine also the other parameters like the
energy scale or the parameter $\alpha$ in the Starobinsky model.

\subsection{Determine the Energy scale and the Parameter $\alpha$}

Starting point is the formula
\begin{equation}
a=a_{0}\cdot\exp(\vartheta)\label{eq:scaling-formula-length}
\end{equation}
with the definition (\ref{eq:shortening-vartheta}) of $\vartheta$.
In \cite{AsselmeyerKrol2014}, we also derived this formula by relating
it to the levels of the tree representing the Casson handle. By using
the shortening $\vartheta$, we obtain 
\[
a=a_{0}\cdot\sum_{n=0}^{\infty}\frac{\vartheta^{n}}{n!}
\]
or the $n$th level will contribute by $\frac{\vartheta^{n}}{n!}$.
To calculate the energy scale, we need the argumentation that the
energy scale as represented by an energy change $\Delta E$ is related
to a time change by $\Delta t\sim h/\Delta E$. Therefore we are enforced
to determine the shortest time change. But this change must agree
with the number of levels in the tree of the Casson handle where the
topology change appears. The Casson handle is designed to produce
a (flat) disk with no self-intersections. As explained above, this
disk will be used to cancel additional self-intersections and at the
end it will lead to the topology change. Therefore we have to ask
how many levels are necessary to get the first disk with no self-intersections.
In \cite{Fre:88} Freeman answered this question: three levels are
needed! Now it seems natural that the shortest time scale will be
assumed to be the Planck time $t_{Planck}$. Then we will get 
\[
\Delta t_{inflation}=\left(1+\vartheta+\frac{\vartheta^{2}}{2}+\frac{\vartheta^{3}}{6}\right)t_{Planck}
\]
for the shortest time interval of the topology change. Finally we
will obtain for the energy scale 
\begin{equation}
\Delta E_{inflation}=\frac{E_{Planck}}{1+\vartheta+\frac{\vartheta^{2}}{2}+\frac{\vartheta^{3}}{6}}\quad.\label{eq:energy-scale-inflation}
\end{equation}
of the inflation. In subsection \ref{sub:A-geometric-interpretation},
we gave a geometrical interpretation of the parameter $\alpha$ as
the radius of a non-contractable core. Following the argumentation
above, then this core has to consist of at least three levels. Furthermore,
$\alpha$ has to be expressed as energy in Planck units. But then
using (\ref{eq:energy-scale-inflation}) we will obtain
\[
\alpha\cdot M_{P}^{-2}=\frac{1}{\left(1+\vartheta+\frac{\vartheta^{2}}{2}+\frac{\vartheta^{3}}{6}\right)}
\]
and as we will see below, $\alpha$ will be of order $10^{-5}$ below
the Planck energy. Finally, the spectral tilt ${\displaystyle n_{s}}$
and the tensor-scalar ratio $r$ can be determined to be
\[
n_{s}=1-\frac{2}{\vartheta+ln(8\pi^{2})}\qquad r=\frac{12}{(\vartheta+ln(8\pi^{2}))^{2}}
\]
with the topological invariant $\vartheta$.

\subsection{Reheating and topology}

Now we will discuss the Einstein-Hilbert action for the sequences
of cobordism $W(Y_{1},Y_{2})\cup_{Y_{2}}W(Y_{2},Y_{3})\cup_{Y_{3}}\cdots$
following our work \cite{AsselmeyerBrans2015}. Let us start with
the (Euclidean) Einstein-Hilbert action functional 
\begin{equation}
S_{EH}(M)=\intop_{M}R\sqrt{g}\: d^{4}x\label{eq:EH-action}
\end{equation}
of the 4-manifold $M$ and fix the Ricci-flat metric $g$ as solution
of the vacuum field equations of the exotic 4-manifold. As discussed
above, we consider a sequences of cobordism 
\[
W(Y_{1},Y_{2})\cup_{Y_{2}}W(Y_{2},Y_{3})\cup_{Y_{3}}\cdots
\]
and one has to consider the Einstein-Hilbert action functional for
every cobordism $W(Y_{n},Y_{n+1})$. In general, for a manifold $M$
with boundary $\partial M=\Sigma$ one has the expression (see \cite{GibHaw1977})
\[
S_{EH}(M)=\intop_{M}R\sqrt{g}\: d^{4}x+\intop_{\Sigma}H\,\sqrt{h}\, d^{3}x
\]
and for the cobordism, one obtains 
\[
S_{EH}(W(Y_{n},Y_{n+1}))=\intop_{W(Y_{n},Y_{n+1})}R\sqrt{g}\, d^{4}x+\intop_{Y_{n+1}}H\,\sqrt{h}\, d^{3}x-\intop_{Y_{n}}H\,\sqrt{h}\, d^{3}x
\]
where $H$ is the mean curvature of the boundary with metric $h$.
In the following we will discuss the boundary term, i.e. we reduce
the problem to the discussion of the action 
\begin{equation}
S_{EH}(\Sigma)=\intop_{\Sigma}H\,\sqrt{h}\, d^{3}x\label{eq:action fermi}
\end{equation}
(see also \cite{Ashtekar08,Ashtekar08a} for this boundary term) along
the boundary $\Sigma$ (a 3-manifold). Now we will show that the action
(\ref{eq:action fermi}) over a 3-manifold $\Sigma$ is equivalent
to the Dirac action of a spinor over $\Sigma$. For completeness we
present the discussion from \cite{AsselmeyerBrans2015}. At first
let us consider the general case of an embedding of a 3-manifold into
a 4-manifold. Let $\iota:\Sigma\hookrightarrow M$ be an embedding
of the 3-manifold $\Sigma$ into the 4-manifold $M$ with the normal
vector $\vec{N}$. A small neighborhood $U_{\epsilon}$ of $\iota(\Sigma)\subset M$
looks like $U_{\epsilon}=\iota(\Sigma)\times[0,\epsilon]$. Furthermore
we identify $\Sigma$ and $\iota(\Sigma)$ ($\iota$ is an embedding).
Every 3-manifold admits a spin structure with a \noun{spin bundle},
i.e. a principal $Spin(3)=SU(2)$ bundle (spin bundle) as a lift of
the frame bundle (principal $SO(3)$ bundle associated to the tangent
bundle). There is a (complex) vector bundle associated to the spin
bundle (by a representation of the spin group), called \noun{spinor
bundle} $S_{\Sigma}$. A section in the spinor bundle is called a
spinor field (or a spinor). In case of a 4-manifold, we have to assume
the existence of a spin structure. But for a manifold like $M$, there
is no restriction, i.e. there is always a spin structure and a spinor
bundle $S_{M}$. In general, the unitary representation of the spin
group in $D$ dimensions is $2^{[D/2]}$-dimensional. From the representational
point of view, a spinor in 4 dimensions is a pair of spinors in dimension
3. Therefore, the spinor bundle $S_{M}$ of the 4-manifold splits
into two sub-bundles $S_{M}^{\pm}$ where one subbundle, say $S_{M}^{+},$
can be related to the spinor bundle $S_{\Sigma}$ of the 3-manifold.
Then the spinor bundles are related by $S_{\Sigma}=\iota^{*}S_{M}^{+}$
with the same relation $\phi=\iota_{*}\Phi$ for the spinors ($\phi\in\Gamma(S_{\Sigma})$
and $\Phi\in\Gamma(S_{M}^{+})$). Let $\nabla_{X}^{M},\nabla_{X}^{\Sigma}$
be the covariant derivatives in the spinor bundles along a vector
field $X$ as section of the bundle $T\Sigma$. Then we have the formula
\begin{equation}
\nabla_{X}^{M}(\Phi)=\nabla_{X}^{\Sigma}\psi-\frac{1}{2}(\nabla_{X}\vec{N})\cdot\vec{N}\cdot\psi\label{eq:covariant-derivative-immersion}
\end{equation}
with the embedding $\phi\mapsto\left(\begin{array}{c}
0\\
\phi
\end{array}\right)=\Phi$ of the spinor spaces from the relation $\phi=\iota_{*}\Phi$. Here
we remark that of course there are two possible embeddings. For later
use we will use the left-handed version. The expression $\nabla_{X}\vec{N}$
is the second fundamental form of the embedding where the trace $tr(\nabla_{X}\vec{N})=2H$
is related to the mean curvature $H$. Then from (\ref{eq:covariant-derivative-immersion})
one obtains the following relation between the corresponding Dirac
operators 
\begin{equation}
D^{M}\Phi=D^{\Sigma}\psi-H\psi\label{eq:relation-Dirac-3D-4D}
\end{equation}
with the Dirac operator $D^{\Sigma}$ on the 3-manifold $\Sigma$.
This relation (as well as (\ref{eq:covariant-derivative-immersion}))
is only true for the small neighborhood $U_{\epsilon}$ where the
normal vector points is parallel to the vector defined by the coordinates
of the interval $[0,\epsilon]$ in $U_{\epsilon}$. In \cite{AsselmeyerRose2012},
we extend the spinor representation of an immersed surface into the
3-space to the immersion of a 3-manifold into a 4-manifold according
to the work in \cite{Friedrich1998}. Then the spinor $\phi$ defines
directly the embedding (via an integral representation) of the 3-manifold.
Then the restricted spinor $\Phi|_{\Sigma}=\phi$ is parallel transported
along the normal vector and $\Phi$ is constant along the normal direction
(reflecting the product structure of $U_{\epsilon}$). But then the
spinor $\Phi$ has to fulfill 
\begin{equation}
D^{M}\Phi=0\label{eq:Dirac-equation-4D}
\end{equation}
in $U_{\epsilon}$ i.e. $\Phi$ is a parallel spinor. Finally we get
\begin{equation}
D^{\Sigma}\psi=H\psi\label{eq:Dirac3D-mean-curvature}
\end{equation}
with the extra condition $|\psi|^{2}=const.$ (see \cite{Friedrich1998}
for the explicit construction of the spinor with $|\psi|^{2}=const.$
from the restriction of $\Phi$). Then we can express the action (\ref{eq:action fermi})
by using (\ref{eq:Dirac3D-mean-curvature}) to obtain 
\begin{equation}
\intop_{\Sigma}H\,\sqrt{h}\, d^{3}x=\intop_{\Sigma}\bar{\psi}\, D^{\Sigma}\psi\,\sqrt{h}d^{3}x\label{eq:relation-mean-curvature-action-to-dirac-action}
\end{equation}
using $|\psi|^{2}=const.$

Above we obtained a relation (\ref{eq:relation-Dirac-3D-4D}) between
a 3-dimensional spinor $\psi$ on the 3-manifold $\Sigma$ fulfilling
a Dirac equation $D^{\Sigma}\psi=H\psi$ (determined by the embedding
$\Sigma\to M$ into a 4-manifold $M$) and a 4-dimensional spinor
$\Phi$ on a 4-manifold $M$ with fixed chirality ($\in\Gamma(S_{M}^{+})$
or $\in\Gamma(S_{M}^{-})$) fulfilling the Dirac equation $D^{M}\Phi=0$
for the 4-dimensional spinor $\Phi$ by using the embedding 
\begin{equation}
\Phi=\left(\begin{array}{c}
0\\
\psi
\end{array}\right)\quad.\label{eq:embedding-spinor-3D-4D}
\end{equation}
In \cite{AsselmeyerBrans2015} we went a step further and discussed
the topology of the 3-manifold leading to a fermion. On general grounds,
one can show that a fermion is given by a knot complement admitting
a hyperbolic structure. The connection between the knot and the particle
properties is currently under investigation. But first calculations
seem to imply that the particular knot is only important for the dynamical
state (like the energy or momentum) but not for charges, flavors etc. 

Now we will reverse the argumentation. Starting with the 4D Dirac
action, the restriction to the boundary is given by
\[
S_{\Sigma}=\intop_{\Sigma}\bar{\psi}\, D^{\Sigma}\psi\,\sqrt{h}d^{3}x-\intop_{\Sigma}H|\psi|^{2}\,\sqrt{h}\, d^{3}x
\]
(where we forget the condition $|\psi|^{2}=const.$). The conformal
transformation (\ref{eq:conformal-trafo-metric}) will also influence
the 3-metric $h$ to get $e^{\phi}h$ so that the last term is given
by
\[
\intop_{\Sigma}H|\psi|^{2}\,\sqrt{h}\, d^{3}x\to\intop_{\Sigma}H\, e^{\phi}\,|\psi|^{2}\,\sqrt{h}\, d^{3}x=S_{WW}
\]
which is an interaction term of the inflaton field $\phi$ with the
fermion field $\psi$. The coupling constant is the mean curvature
$H$ of $\Sigma$ which is constant (by using the hyperbolic structure).
In \cite{AsselmeyerBrans2015}, the extension of this action to the
spacetime $M$ was also discussed for the Dirac operator. If we fix
$H$ as coupling parameter, the extension of $S_{WW}$ to 4D can be
done by using the embedding (\ref{eq:embedding-spinor-3D-4D}). Then
we will get the Lagrangian of the matter-scalar field coupling
\[
\mathcal{L}=\bar{\Phi}\, D^{M}\Phi+|H|e^{\phi}|\Phi|^{2}
\]
and we have to determine the coupling $|H|$ now. The argumentation
follows from the calculation of $\vartheta$ in the previous subsections.
The mean curvature is constant by the hyperbolic geometry. Furthermore,
the fermion $\Phi$ couples geometrically to one level (1-level) of
the Casson handle. Therefore using the scaling formula (\ref{eq:scaling-formula-length}),
we will get 
\[
|H|=\exp\left(-\frac{3}{CS(1-level)}\right)
\]
by using the reference to the Planck scale. Now we have to discuss
the value for the Chern-Simons invariant of 1-level. The corresponding
3-manifold to 1-level is given by complements of the Whitehead link
$Wh$ with $CS(S^{3}\setminus Wh)=\frac{1}{4}\bmod1$ (see the Fig.
in \cite{AsselmeyerKrol2018a}) so that
\begin{equation}
|H|=e^{-12}\approx0.6\cdot10^{-5}\label{eq:coupling-constant}
\end{equation}
gives the coupling between the scalar field $\phi$ and fermion $\Phi$.
In \cite{inflation-reheating-Ellis2015}, they gave an upper bound
of order $10^{-5}$ for the coupling together with an e-fold $N\approx51.7$
to get the right reheating temperature. Later, we will present a particular
model with this number of e-folds. Here, we remark that the coupling
above does not depend on the topology change. It is a universal coupling
between the scalar field (as substitute for the Casson handle) and
the fermion field (as substitute for the knot complement). Therefore,
we obtained a natural model for the coupling of matter to the inflaton
field $\phi$ after inflation.

\section{A realistic example}

The distinguished feature of differential topology of manifolds in
dimension 4 is the existence of open 4-manifolds carrying a plenty
of non-diffeomorphic smooth structures. In the cosmological model
presented here, the special role is played by the topologically simplest
4-manifold, i.e. $\mathbb{R}^{4}$, which carries a continuum of infinitely
many different smoothness structures. Each of them except one, the
standard $\mathbb{R}^{4}$, is called \emph{exotic} $R^{4}$. All
exotic $R^{4}$ are Riemannian smooth open 4- manifolds homeomorphic
to $\mathbb{R}^{4}$ but non-diffeomorphic to the standard smooth
$\mathbb{R}^{4}$. The standard smoothness is distinguished by the
requirement that the topological product $\mathbb{R}\times\mathbb{R}^{3}$
is a smooth product. There exists only one (up to diffeomorphisms)
smoothing, the standard $\mathbb{R}^{4}$, where the product above
is smooth. There two types of exotic $\mathbb{R}^{4}$: small exotic
$\mathbb{R}^{4}$ can be embedded into the standard $S^{4}$ whereas
large exotic $\mathbb{R}^{4}$ cannot. In the following, an exotic
$\mathbb{R}^{4}$, presumably small if not stated differently, will
be denoted as $R^{4}$. In cosmology, one usually considers the topology
$S^{3}\times\mathbb{R}$ for the spacetime. But by using the simple
topological relations $\mathbb{R}^{4}\setminus D^{4}=S^{3}\times\mathbb{R}$
or $\mathbb{R}^{4}\setminus\left\{ 0\right\} =S^{3}\times\mathbb{R}$,
one obtains also an exotic $S^{3}\times\mathbb{R}$ from every exotic
$R^{4}$. In the following we will denote the exotic $S^{3}\times\mathbb{R}$
by $S^{3}\times_{\theta}\mathbb{R}$ to indicate the important fact
that there is no global splitting of $S^{3}\times_{\theta}\mathbb{R}$
or it is not globally hyperbolic. This fact has a tremendous impact
on cosmology and therefore we will consider our main hypothesis:\\
 \emph{MainHypo: The spacetime, seen as smooth four-dimensional manifold,
admits an exotic smoothness structure.}

\subsection{Introduction of the model}

The main hypothesis above has the following consequences (see \cite{AsselmeyerKrol2018a}): 
\begin{itemize}
\item Any $R^{4}$ has necessarily non-vanishing Riemann curvature. Also
the $S^{3}\times_{\theta}\mathbb{R}$ has a non-vanishing curvature. 
\item Inside of $R^{4}$, there is a compact 4-dimensional submanifold $K\subset R^{4}$,
which is not surrounded by a smoothly embedded 3-sphere. Then there
is a chain of 3-submanifolds of $R^{4}$ $Y_{1}\to\cdots\to Y_{\infty}$
and the corresponding infinite chain of cobordisms 
\[
End(R^{4})=W(Y_{1},Y_{2})\cup_{Y_{2}}W(Y_{2},Y_{3})\cup\cdots
\]
where $W(Y_{k},Y_{k+1})$ denotes the cobordism between $Y_{k}$ and
$Y_{k+1}$ so that $R^{4}=K\cup_{Y_{1}}End(R^{4})$ where $\partial K=Y_{1}$.
Furthermore one has $End(R^{4})\subset S^{3}\times_{\theta}\mathbb{R}$. 
\item $R^{4}$ and $S^{3}\times_{\theta}\mathbb{R}$ embeds into the standard
$\mathbb{R}^{4}$ or $S^{4}$ but also in some other complicated 4-manifolds.
The construction of $R^{4}$ gives us a natural smooth embedding into
the compact 4-manifold $E(2)\#\overline{\mathbb{C}P^{2}}$ (with the
K3 surface $E(2)$) (see \cite{BizGom:96}). 
\end{itemize}
But every subset $K'$, $K'\subset K\subset R^{4}$, is surrounded
by a 3-sphere. This fact is the starting point of our model. Now we
choose Planck-size 3-sphere $S^{3}$ inside of the compact subset
$K\subset R^{4}$. This is the initial point where our cosmos starts
to evolve. By the construction of $R^{4}$, as mentioned above, there
exists the homology 3-sphere 
\[
\Sigma(2,5,7)=\left\{ (x,y,z)\in\mathbb{C}^{3}|\, x^{2}+y^{5}+z^{7}=0\,,\:|x|^{2}+|y|^{2}+|z|^{2}=1\right\} 
\]
 inside of $K$ which is the boundary of the Akbulut cork for $E(2)\#\overline{\mathbb{C}P^{2}}$.
(see chapter 9, \cite{GomSti:1999}). If $S^{3}$ is the starting
point of the cosmos as above, then $S^{3}\subset\Sigma(2,5,7)$. But
then we will obtain the first topological transition 
\[
S^{3}\to\Sigma(2,5,7)
\]
inside $R^{4}$. The construction of $R^{4}$ was based on the topological
structure of $E(2)$ (the K3 surface). $E(2)$ splits topologically
into a 4-manifold $|E_{8}\oplus E_{8}|$ with intersection form $E_{8}\oplus E_{8}$
(see \cite{GomSti:1999}) and the sum of three copies of $S^{2}\times S^{2}$.
In this topological splitting 
\begin{equation}
|E_{8}\oplus E_{8}|\times\underbrace{\left(S^{2}\times S^{2}\right)\times\left(S^{2}\times S^{2}\right)\times\left(S^{2}\times S^{2}\right)}_{3\left(S^{2}\times S^{2}\right)}\label{eq:splitting-K3}
\end{equation}
the 4-manifold $|E_{8}\oplus E_{8}|$ has a boundary which is the
sum of two Poincar{é} spheres $P\#P$. Here we used the fact that
a smooth 4-manifold with intersection form $E_{8}$ must have a boundary
(which is the Poincar{é} sphere $P$), otherwise it would contradict
the Donaldson's theorem. Then any closed version of $|E_{8}\oplus E_{8}|$
does not exist and this fact is the reason for the existence of $R^{4}$.
To express it differently, the $R^{4}$ lies between this 3-manifold
$\Sigma(2,5,7)$ and the sum of two Poincar{é} spheres $P\#P$.
We analyzed this spacetime in\cite{AsselmeyerKrol2012}. It is interesting
to note that \emph{the number of $S^{2}\times S^{2}$ components must
be three or more otherwise the corresponding spacetime is not smooth!}

Therefore we have two topological transitions resulting from the embedding
into $E(2)\#\overline{\mathbb{C}P^{2}}$ 
\[
S^{3}\stackrel{cork}{\longrightarrow}\Sigma(2,5,7)\stackrel{gluing}{\longrightarrow}P\#P\,.
\]
These two topological transition are the main idea of our model.

\subsection{Consequences for the inflation}

In this section we will show how the change of the energy scale is
driven by the topological transitions
\[
S^{3}\stackrel{cork}{\longrightarrow}\Sigma(2,5,7)\stackrel{gluing}{\longrightarrow}P\#P\,.
\]
Both transitions have different topological descriptions. The first
transition $S^{3}\to\Sigma(2,5,7)$ can be realized by a smooth cobordism.
To realize this cobordism, one has to a add a 1-/2-handle pair together
with one relation. It is a characteristic property of the 4-dimensional
spacetime that this adding of a handle pair will also produces extra
intersections between the handles. For that purpose we need a structure,
which is an infinite tree of self-intersecting disks, also known as
Casson handle. In \cite{AsselmeyerKrol2018a} we described this situation
extensively. If one assumes a Planck-size $(L_{P})$ 3-sphere at the
Big Bang then the scale $a$ of $\Sigma(2,5,7)$ changes like
\[
a=L_{P}\cdot\exp\left(\frac{3}{2\cdot CS(\Sigma(2,5,7))}\right)
\]
with the Chern-Simons invariant and $\vartheta$ 
\[
CS(\Sigma(2,5,7))=\frac{9}{4\cdot(2\cdot5\cdot7)}=\frac{9}{280}\qquad\vartheta=\frac{140}{3}
\]
and the Planck scale of order $10^{-34}m$ changes to $10^{-15}m$.
Obviously, this transition has an exponential or inflationary behavior.
Surprisingly, the number of e-folds can be explicitly calculated (see
\cite{AsselmeyerKrol2018b}) by the formula (\ref{eq:number-of-e-folds})
to be
\begin{equation}
N=\frac{3}{2\cdot CS(\Sigma(2,5,7))}+ln(8\pi^{2})\approx51\label{eq:e-fold-first}
\end{equation}
and we also obtain the energy and time scale of this transition (see
\cite{AsselmeyerKrol2018b})
\begin{equation}
E_{GUT}=\frac{E_{P}}{1+\vartheta+\frac{\vartheta^{2}}{2}+\frac{\vartheta^{3}}{6}}\approx10^{15}GeV\quad t=t_{P}\left(1+\vartheta+\frac{\vartheta^{2}}{2}+\frac{\vartheta^{3}}{6}\right)\approx10^{-39}s\label{eq:GUT-energy-scale}
\end{equation}
right at the conjectured GUT scale ($E_{P},t_{P}$ Planck energy and
time, respectively). Above, we showed that this transition is described
But then, the dimension-less free parameter $\alpha\cdot M_{P}^{-2}$
as well spectral tilt ${\displaystyle n_{s}}$ and the tensor-scalar
ratio $r$ can be determined to be
\[
\alpha\cdot M_{P}^{-2}=1+N+\frac{N^{2}}{2}+\frac{N^{3}}{6}\approx10^{-5},\: n_{s}\approx0.961\,,\: r\approx0.0046
\]
using (\ref{eq:e-fold-first}) which is in good agreement with current
measurements. As discussed in the previous section, this number of
e-folds is compatible with the coupling constant (\ref{eq:coupling-constant})
in the reheating process.

\section{Conclusion}

In this paper we presented a complete topological picture to describe
inflation. We started with a general formalism of topology change
using the cobordism concept.This description lead naturally to the
introduction of a scalar field $\phi$. The potential $V(\phi)$ is
given by the squared derivative of the Morse function. By using Cerf
theory, we obtained two possible models: chaotic inflation $V(\phi)\sim\phi^{2}$
and topological inflation $V(\phi)\sim(\phi^{2}-1)^{2}$. Both models
were ruled out by the Planck mission. But in particular for 4-dimensional
spacetime, there is another possibility which leads naturally to Starobinsky
inflation. According to this model, an inflationary phase in the cosmic
evolution is caused by the exotic smoothness structure of our spacetime.
The exotic smoothness structure is constructed by a hyperbolic homology
3-sphere $\Sigma$. The exponential expansion has its origin in the
hyperbolic structure of the spacetime. This expansion is determined
by a single parameter $\vartheta$, the fraction of two topological
invariants for the hyperbolic homology 3-sphere: the volume and the
Chern-Simons invariant. Furthermore, we were able to calculate the
number of e-folds which is by Mostow-Prasad rigidity a topological
invariant. With the help of the invariant, we were also able to determine
all parameters of the Starobinsky model like $\alpha$, energy scale,
e-folds and the coupling constant for the reheating process. One question
remains: But what is about the inflation without quantum effects?
Fortunately, there is growing evidence that the differential structures
constructed above (i.e. exotic smoothness in dimension 4) is directly
related to quantum gravitational effects \cite{Ass2010,Duston2010,AsselmeyerMaluga2016}.
Maybe we touch only the tip of the iceberg. 
\begin{acknowledgments}
At first we want to express our gratitude to C.H. Brans and R. Gompf
for numerous discussions. This publication was made possible through
the support of a grant from the John Templeton Foundation (Grant No.
60671).
\end{acknowledgments}
\appendix

\section{Connected and boundary-connected sum of manifolds\label{sec:Connected-and-boundary-connected}}

Now we will define the connected sum $\#$ and the boundary connected
sum $\natural$ of manifolds. Let $M,N$ be two $n$-manifolds with
boundaries $\partial M,\partial N$. The \emph{connected sum} $M\#N$
is the procedure of cutting out a disk $D^{n}$ from the interior
$int(M)\setminus D^{n}$ and $int(N)\setminus D^{n}$ with the boundaries
$S^{n-1}\sqcup\partial M$ and $S^{n-1}\sqcup\partial N$, respectively,
and gluing them together along the common boundary component $S^{n-1}$.
The boundary $\partial(M\#N)=\partial M\sqcup\partial N$ is the disjoint
sum of the boundaries $\partial M,\partial N$. The \emph{boundary
connected sum} $M\natural N$ is the procedure of cutting out a disk
$D^{n-1}$ from the boundary $\partial M\setminus D^{n-1}$ and $\partial N\setminus D^{n-1}$
and gluing them together along $S^{n-2}$ of the boundary. Then the
boundary of this sum $M\natural N$ is the connected sum $\partial(M\natural N)=\partial M\#\partial N$
of the boundaries $\partial M,\partial N$.

\section{Casson Handles}

Let us start with the basic construction of the Casson handle $CH$.
Let $M$ be a smooth, compact, simple-connected 4-manifold and $f:D^{2}\to M$
a (codimension-2) mapping. By using diffeomorphisms of $D^{2}$ and
$M$, one can deform the mapping $f$ to get an immersion (i.e. injective
differential) generically with only double points (i.e. $\#|f^{-1}(f(x))|=2$)
as singularities \cite{GolGui:73}. But to incorporate the generic
location of the disk, one is rather interesting in the mapping of
a 2-handle $D^{2}\times D^{2}$ induced by $f\times id:D^{2}\times D^{2}\to M$
from $f$. Then every double point (or self-intersection) of $f(D^{2})$
leads to self-plumbings of the 2-handle $D^{2}\times D^{2}$. A self-plumbing
is an identification of $D_{0}^{2}\times D^{2}$ with $D_{1}^{2}\times D^{2}$
where $D_{0}^{2},D_{1}^{2}\subset D^{2}$ are disjoint sub-disks of
the first factor disk%
\footnote{In complex coordinates the plumbing may be written as $(z,w)\mapsto(w,z)$
or $(z,w)\mapsto(\bar{w},\bar{z})$ creating either a positive or
negative (respectively) double point on the disk $D^{2}\times0$ (the
core).%
}. Consider the pair $(D^{2}\times D^{2},\partial D^{2}\times D^{2})$
and produce finitely many self-plumbings away from the attaching region
$\partial D^{2}\times D^{2}$ to get a kinky handle $(k,\partial^{-}k)$
where $\partial^{-}k$ denotes the attaching region of the kinky handle.
A kinky handle $(k,\partial^{-}k)$ is a one-stage tower $(T_{1},\partial^{-}T_{1})$
and an $(n+1)$-stage tower $(T_{n+1},\partial^{-}T_{n+1})$ is an
$n$-stage tower union kinky handles $\bigcup_{\ell=1}^{n}(T_{\ell},\partial^{-}T_{\ell})$
where two towers are attached along $\partial^{-}T_{\ell}$. Let $T_{n}^{-}$
be $(\mbox{interior}T_{n})\cup\partial^{-}T_{n}$ and the Casson handle
\[
CH=\bigcup_{\ell=0}T_{\ell}^{-}
\]
is the union of towers (with direct limit topology induced from the
inclusions $T_{n}\hookrightarrow T_{n+1}$).

The main idea of the construction above is very simple: an immersed
disk (disk with self-intersections) can be deformed into an embedded
disk (disk without self-intersections) by sliding one part of the
disk along another (embedded) disk to kill the self-intersections.
Unfortunately the other disk can be immersed only. But the immersion
can be deformed to an embedding by a disk again etc. In the limit
of this process one ''shifts the self-intersections into infinity''
and obtains%
\footnote{In the proof of Freedman \cite{Fre:82}, the main complications come
from the lack of control about this process. %
} the standard open 2-handle $(D^{2}\times\mathbb{R}^{2},\partial D^{2}\times\mathbb{R}^{2})$.

A Casson handle is specified up to (orientation preserving) diffeomorphism
(of pairs) by a labeled finitely-branching tree with base-point {*},
having all edge paths infinitely extendable away from {*}. Each edge
should be given a label $+$ or $-$. Here is the construction: tree
$\to CH$. Each vertex corresponds to a kinky handle; the self-plumbing
number of that kinky handle equals the number of branches leaving
the vertex. The sign on each branch corresponds to the sign of the
associated self plumbing. The whole process generates a tree with
infinite many levels. In principle, every tree with a finite number
of branches per level realizes a corresponding Casson handle. Technically
speaking, each building block of a Casson handle, the ``kinky''
handle with $n$ kinks%
\footnote{The number of end-connected sums is exactly the number of self intersections
of the immersed two handle.%
}, is diffeomorphic to the $n-$times boundary-connected sum $\natural_{n}(S^{1}\times D^{3})$
(see appendix \ref{sec:Connected-and-boundary-connected}) with two
attaching regions. One region is a tubular neighborhood of band sums
of Whitehead links connected with the previous block. The other region
is a disjoint union of the standard open subsets $S^{1}\times D^{2}$
in $\#_{n}S^{1}\times S^{2}=\partial(\natural_{n}S^{1}\times D^{3})$
(this is connected with the next block).

\section{Hyperbolic 3-/4-Manifolds and Mostow-Prasad rigidity\label{sec:Appendix-Hyperbolic-Mostow}}

In short, Mostow\textendash{}Prasad rigidity theorem states that the
geometry of a complete, finite-volume hyperbolic manifold of dimension
greater than two is uniquely determined by the fundamental group.
The corresponding theorem was proven by Mostow for closed manifolds
and extended by Prasad for finite-volume manifolds with boundary.
In dimension 3, there is also an extension for non-compact manifolds
also called ending lamination theorem. It states that hyperbolic 3-manifolds
with finitely generated fundamental groups are determined by their
topology together with invariants of the ends admitting a kind of
foliation at surfaces in the end. The end of a 3-manifolds has always
the form $S\times[0,1)$ with the compact surfaces $S$. Then a lamination
on the surface $S$ is a closed subset of $S$ that is written as
the disjoint union of geodesics of $S$.

A general formulation of the Mostow-Prasad rigidity theorem is:\\
 Let $M,N$ be compact hyperbolic $n-$manifolds with $n\geq3$. Assume
that $M$ and $N$ have isomorphic fundamental groups. Then the isomorphism
of fundamental groups is induced by a unique isometry.\\
 An important corollary states that \emph{geometric invariants are
topological invariants}. The Mostow-Prasad rigidity theorem has special
formulations for dimension 3 and 4. Both manifolds $M,N$ have to
be homotopy-equivalent and every homotopy-equivalence induces an isometry.
In dimension 3, the homotopy-equivalence of a 3-manifold of non-positive
sectional curvature implies a homeomorphism (a direct consequence
of the geometrization theorem, the exception are only the lens spaces)
and a diffeomorphism (see Moise \cite{Moi:52}). In dimension 4, compact
homotopy-equivalent simply-connected 4-manifolds are homeomorphic
(see Freedman \cite{Fre:82}). This result can be extended to a large
class of compact non-simply connected 4-manifolds (having a good fundamental
group), see \cite{FreQui:90}. Therefore, if a 3- or 4-manifold admits
a hyperbolic structure then this structure is unique up to isometry
and all geometric invariants are topological invariants among them
the volume and the curvature.

Then a hyperbolic 3-manifold $M^{3}$ is given by the quotient space
$\mathbb{H}^{3}/\Gamma$ where $\Gamma\subset Isom(\mathbb{H}^{3})=SO(3,1)$
is a discrete subgroup (Kleinian group) so that $\Gamma\simeq\pi_{1}(M^{3})$.
A hyperbolic structure is a homomorphism $\pi_{1}(M^{3})\to SO(3,1)$
up to conjugacy (inducing the isometry). The analogous result holds
for the hyperbolic 4-manifold which can be written as quotient $\mathbb{H}^{4}/\pi_{1}(M^{4})$.

Let $X^{4}$ be a compact hyperbolic 4-manifold with metric $g_{0}$
and let $M^{4}$ be a compact manifold together with a smooth map
$f:M\to X$. As shown in \cite{BessonCourtoisGallot:1995} or in the
survey \cite{BessonCourtoisGallot:1996} (Main Theorem 1.1), the volumes
of $X,M$ are related 
\[
Vol_{r}(M)\geq deg(f)Vol(X,g_{0})
\]
where $deg(f)$ denotes the degree of $f$. If equality holds, and
if the infimum of the relation is achieved by some metric $g$, then
$(M,g)$ is an isometric Riemannian covering of $(X,g_{0})$ with
covering map $M\to X$ homotopic to $f$. In particular, if $f$ is
the identity map $X\to X$ (having degree $deg(f)=1$) then it implies
that $g_{0}$ is the only Einstein metric on $X$ up to rescalings
and diffeomorphisms.

\section{Models of Hyperbolic geometry\label{sec:Appendix-models-hyp-geom}}

In the following we will describe two main models of hyperbolic geometry
which were used in this paper. For simplicity we will concentrate
on the two-dimensional versions.

The Poincare disk model also called the conformal disk model, is a
model of 2-dimensional hyperbolic geometry in which the points of
the geometry are inside the unit disk, and the straight lines consist
of all segments of circles contained within that disk that are orthogonal
to the boundary of the disk, plus all diameters of the disk. The metric
in this model is given by 
\[
ds^{2}=\frac{dx^{2}+dy^{2}}{(1-(x^{2}+y^{2}))^{2}}
\]
which can be transformed to expression (\ref{eq:metric-hyp-4-ball})
by a radial coordinate transformation. In this model, the hyperbolic
geometry is confined to the unit disk, where the boundary represents
the 'sphere at infinity'.

The Poincare half-plane model is the upper half-plane, denoted by
$\mathbb{H}^{2}=\left\{ (x,y)\:|\: y>0,\: x,y\in\mathbb{R}\right\} $,
together with a metric, the Poincare metric, 
\[
ds^{2}=\frac{dx^{2}+dy^{2}}{y^{2}}
\]
(see (\ref{eq:hyp-half-plane})) that makes it a model of two-dimensional
hyperbolic geometry. Here the line $y=0$ represents the infinity
(so-called ideal points).

Both models are isometric to each other. A point $(x,y)$ in the disk
model maps to the point 
\[
\left(\frac{2x}{x^{2}+(1-y)^{2}},\frac{1-x^{2}-y^{2}}{x^{2}+(1-y)^{2}}\right)
\]
in the half-plane model conversely a point $(x,y)$ in the half-plane
model maps to the point 
\[
\left(\frac{2x}{x^{2}+(1+y)^{2}},\frac{x^{2}+y^{2}-1}{x^{2}+(1+y)^{2}}\right)
\]
in the disk model. This transform is known as Cayley transform.


\end{document}